\documentclass[twocolumn, 10pt, journal]{IEEEtran}
\IEEEoverridecommandlockouts
\usepackage{lineno}
\usepackage{subfigure}

\usepackage{graphicx}
\usepackage{subfigure}
\usepackage{amsmath}
\usepackage{cite}
\usepackage{float}
\usepackage{lettrine}
\usepackage{soul}
\usepackage{color}
\usepackage[table]{xcolor}
\usepackage{diagbox}
\usepackage{multirow}
\usepackage{array}
\usepackage[ruled]{algorithm2e}
\sethlcolor{yellow}

\newcolumntype{M}[1]{>{\centering\arraybackslash}m{#1}}
\newcolumntype{N}[1]{>{\raggedright\arraybackslash}m{#1}}
\newcommand{\eat}[1]

\usepackage{booktabs}

\begin{document}

\title{Enabling Cyberattack-Resilient Load Forecasting through Adversarial Machine Learning
\thanks{This work was supported in part by the National Science Foundation under Grants ECCS-1611095, CNS-1647209 and ECCS-1831811, in part by the Department of Energy Cybersecurity for Energy Delivery Systems (CEDS), and in part by the Office of the Provost, University of Connecticut.}
\thanks{Z. Tang and P. Zhang are with the Department of Electrical and Computer Engineering, Stony Brook University, Stony Brook, NY 11794, USA (e-mail: p.zhang@stonybrook.edu).}
\thanks{J. Jiao and J. Yan are with the Department of Statistics, University of Connecticut, Storrs, CT 06269, USA.}
\thanks{M. Yue is with Sustainable Energy Technologies Department, Brookhaven National Laboratory, Upton, NY 11973, USA.}
\thanks{C. Chen is with Energy Systems Division, Argonne National Laboratory, Lemont, IL 60439, USA.}
}

\author{\IEEEauthorblockN{Zefan Tang,~\IEEEmembership{Student Member,~IEEE,} Jieying Jiao,~\IEEEmembership{Student Member,~IEEE,} Peng Zhang,~\IEEEmembership{Senior Member,~IEEE,} Meng Yue,~\IEEEmembership{Member,~IEEE,} Chen Chen,~\IEEEmembership{Member,~IEEE,}  Jun Yan}}

\maketitle

\begin{abstract}
In the face of an increasingly broad cyberattack surface, cyberattack-resilient load forecasting for electric utilities is both more necessary and more challenging than ever. In this paper, we propose an adversarial machine learning (AML) approach, which can respond to a wide range of attack behaviors without detecting outliers. It strikes a balance between enhancing a system's robustness against cyberattacks and maintaining a reasonable degree of forecasting accuracy when there is no attack. Attack models and configurations for the adversarial training were selected and evaluated to achieve the desired level of performance in a simulation study. The results validate the effectiveness and excellent performance of the proposed method.
\end{abstract}


\section{Introduction}
\IEEEPARstart{F}{ORECASTING} the electricity load is of paramount importance in the operation, planning and marketing of power systems. It is greatly needed at both transmission and distribution levels for various tasks such as hydro scheduling, hydro-thermal coordination, unit commitment, economic dispatch, automatic generation control, load flow analysis, power purchase, and load switching~\cite{gross1987short,tang2018extreme}. Traditionally, utilities were mainly concerned with the timeliness and accuracy of the predictions. Issues were mainly focused from the economic perspective. For instance, a better load forecasting enables utilities to properly assign online and offline reserves since the cost of reserves is very high~\cite{doherty2005new}. Very little attention was paid to cyber security issues.

However, with the increasing deployment of smart grid technologies like sensing, digital control, and communication infrastructure, the data needed as input for forecasting models can be compromised by a cyber adversary through various means. For instance, real-time forecasting data significantly rely on power grids' communication, control, and computing infrastructures, as well as their hardware facilities, all of which are vulnerable to attacks~\cite{sridhar2012cyber,gusrialdi2019smart}. The aggregation of data usually requires diverse data sources, which also gives rise to a wide attack surface~\cite{wang2019online,lu2018coupled}. Moreover, the cryptographic algorithms used to preserve the aggregated data can be decrypted if the attackers have strong capabilities or have enough time to improve their abilities~\cite{sun2018cyber}. Further, the long duration of data preservation requires the migration of data, and the migration process also poses security challenges~\cite{rajesh2019securing}.


A common approach to cyberattack-resilient load forecasting is to remove malicious data through the use of anomaly detection techniques~\cite{cui2019machine}. Some methods are based on descriptive analytics, which identify point, contextual, and collective anomalies such as abnormal patterns~\cite{yue2019descriptive,guo2012detecting}. Other methods are model-based and compare predicted values with observed ones~\cite{xie2016gefcom2014,jian2018real}. Although anomaly detection is an important step in cleaning the input data, some anomalies may remain undetected, and false alarms may also be raised, which still need to be dealt with by the forecasting model~\cite{giani2013smart,boroojeni2017bad,wang2018review,chen2019exploiting}. It is necessary to reduce the sensitivity of the forecasting model to malicious input data and, thus, mitigate the impact of unidentified cyberattacks.

Many robust load forecasting methods have been devised by down-weighting the observations that are more likely to be anomalies. For instance, the weight functions of Huber's seminal work~\cite{huber1973robust} are widely used. \cite{gelper2010robust} used robust versions of the exponential and Holt-Winters smoothing methods. Robust ensemble approaches such as robust functional principal component analysis have also been applied to load forecasting \cite{hyndman2007robust,alobaidi2018robust,vilar2016using}. However, most of the existing robust methods are only concerned with outliers, namely, the extremely high/low observations. More sophisticated and better disguised attacks may not be remedied by the existing robust methods.

Adversarial machine learning (AML) is a recent technique that promises to enhance the robustness of machine learning (ML) based forecasting methods against cyberattacks. The traditional practice in training ML models is to use clean data only, but this means that the output will be erroneous if the input data become contaminated. AML trains the model with both clean data and malicious data generated by the defender~\cite{cai2018curriculum}. The adversarially trained models are then robust to the attacks used in the adversarial training~\cite{tramer2017ensemble,tan2018generative}. However, most existing works on AML are on image recognition, where there is little ambiguity in the output regardless of the adversarial input~\cite{shaham2018understanding}. In the context of cyberattack-resilient load forecasting, AML approaches to image recognition cannot be directly applied. It was unclear how to set attack models and their parameters during the adversarial training stage.

In this paper, we investigate the feasibility of using AML for an artificial neural network (ANN)-based load forecasting. A practical study is constructed using the publicly available data from the Global Energy Forecasting Competition 2012 (GEFCom2012)~\cite{hong2014global,luo2018benchmarking}. This is the first application of AML to cyberattack-resilient load forecasting. The main contributions of this paper are fourfold:
\begin{itemize}
    \item The impacts of different cyberattacks on the traditional ANN-based load forecasting are evaluated.
    \item An accelerated adversarial training is developed for AML-based load forecasting. It maintains the robustness, and at the same time is efficient and easy to employ.
    \item A strategic framework for AML-based load forecasting is established. It strikes a balance between two objectives: ensuring robust performance in forecasting error under various attacks and under no attack.
    \item The selection of attack models and configurations for the adversarial training is extensively investigated in a simulation study. It provides valuable insights of how to set attack models and their parameters during the adversarial training stage.
\end{itemize}

The rest of this paper is organized as follows: The cyber security issues with ANN-based load forecasting are reviewed in Section~\ref{sec:cyber}. Section~\ref{sec:aml} discusses how AML is adapted to load forecasting and provides specifics on the strategies used to select adversary models in the training stage. The results of our numerical investigation are reported in Section~\ref{sec:result}. Section~\ref{sec:conclusion} concludes the paper.

\section{ANN Load Forecasting Under Cyberattacks}
\label{sec:cyber}
This section describes the data, the ANN load forecasting model, and the attack models. It also investigates the performance of the ANN model under various types of attacks.
\vspace{-4pt}
\subsection{Data Description}
The publicly available GEFCom2012 dataset has been widely used as a testbed in the load forecasting community. For instance, it has been used to evaluate the forecasting accuracy of various short-term load forecasting methods~\cite{luo2018benchmarking,nowotarski2016improving,lloyd2014gefcom2012}. Specifically, this dataset consists of 4.5 years of hourly loads and temperatures from 20 zones. In this paper, we took three years (2004-2006) of load data from the first zone for our empirical study. Two years' data spanning 2004 and 2005 were used as the training data, and the 2006 data were used as the testing data. For more details of the data used in this study, readers are referred to \cite{hong2014global}.

\subsection{ANN Load Forecasting Model}

Machine learning approaches have increasingly attracted attention in forecasting electricity load because of their flexibility and ease of use~\cite{liang2019short}. A typical ANN consists of a collection of neurons, each of which receives multiple inputs, processes them internally, and outputs a response. Different weights are allocated during the linear combination of the inputs, and the result is further handled via a non-linear activation function, e.g., a sigmoid function.
A multi-layered ANN consists of an input layer, one or more hidden layers, and an output layer. The network propagates the values from the input layer through the hidden layer(s) to the output layer, where a squared error loss can be obtained
\begin{equation}
L(\hat{z}_m, z_m)=\frac{1}{2}(\hat{z}_m-z_m)^2,
\label{equ:sel}
\end{equation}
where $m$ denotes the $m^{th}$ pair of the input and output, $\hat{z}_m$ is the network output after using the activation function, $z_m$ is the actual value, and $L(\hat{z}_m, z_m)$ is the squared error loss between $\hat{z}_m$ and $z_m$. A loss function can thus be computed as
\begin{equation}
E=\frac{1}{N_r}\sum_{m=1}^{N_r}L(\hat{z}_m, z_m),
\label{equ:loss}
\end{equation}
where $E$ is the loss function, and $N_r$ is the number of the input and output pairs in the training dataset.

During the learning stage, the network is updated iteratively by changing its weights until $E$ is minimized or lower than a threshold. With different weights, the trained ANN model will have different performances.
Many optimization methods have been devised to achieve this goal. One commonly used approach is the back-propagation, which uses the stochastic gradient descent (SGD) method to estimate the gradients of $E$ with respect to the weights. Each weight $w$ is updated as follows:
\begin{equation}
w_{j+1}=w_j-\frac{\eta}{N_r}\sum_{m=1}^{N_r} \nabla_w L(\hat{z}_{m,j}, z_{m}),
\label{equ:weight}
\end{equation}
where $j$ denotes the $j^{th}$ iteration, $\hat{z}_{m,j}$ is the $m^{th}$ output value at the $j^{th}$ iteration, and $\eta$ is the learning rate which controls the speed of the training. The initial weights are set randomly.

\begin{figure}[t]
\centering
\includegraphics[width=0.48\textwidth]{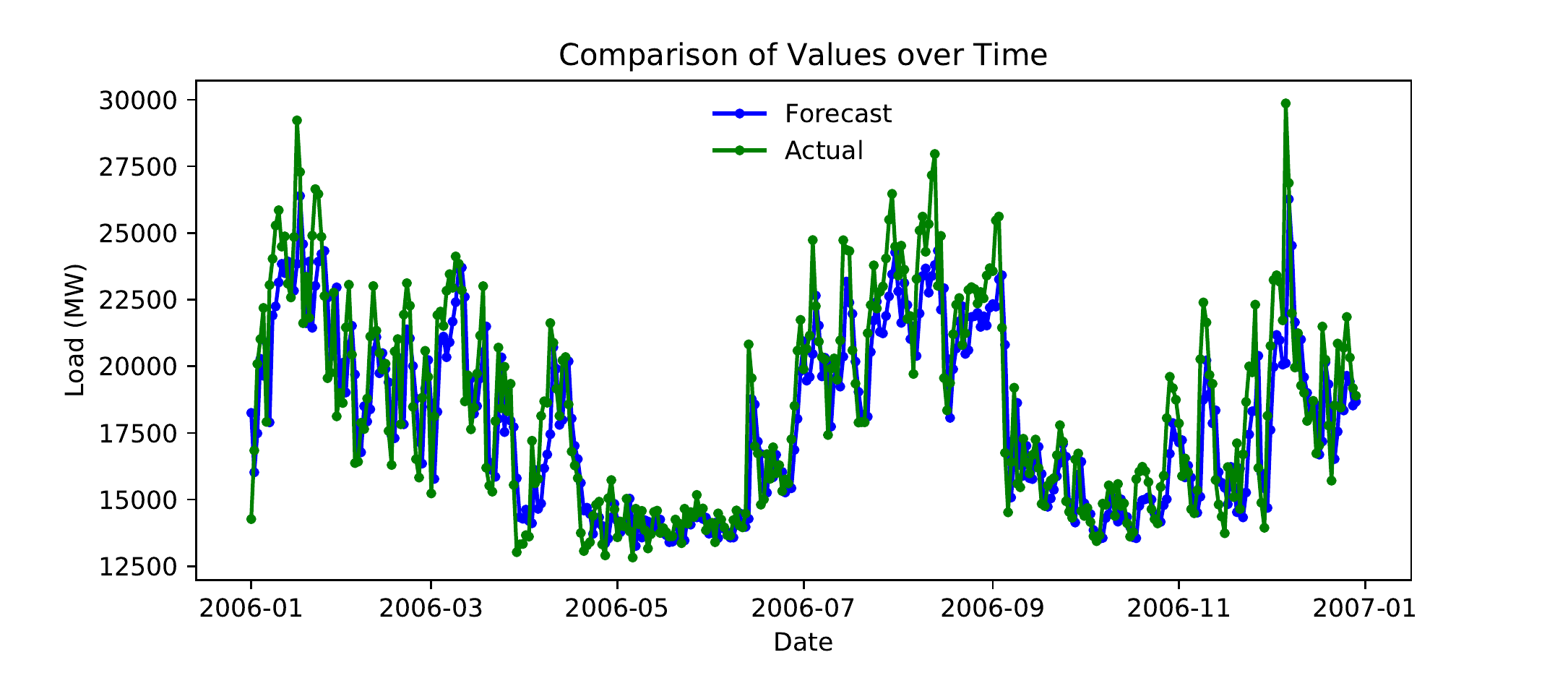}
\vspace{-8pt}
\caption{\label{fig:forecast}Comparison results of ANN-based load forecasting and actual values on the testing data throughout 2006.}
\vspace{-4pt}
\end{figure}

In this study, without loss of generality, the hourly loads on the past seven days are used as the input to the ANN, and the output of the network is the average load on the next day. This forecasting provides important information so that the wholesale market can better understand the loads in specific regions or load zones~\cite{wholesaleload}. To improve the effectiveness of the learning process, the input and output training data and the input testing data are scaled through the use of natural logarithm before the learning stage. In this study, the ANN model has three layers with 50 neurons in the hidden layer. The relationship between the input and output is mapped, learned, and stored into the weights via the two-year training dataset. The performance of the trained ANN is examined with the one-year testing data, as shown in Fig.~\ref{fig:forecast}.

To evaluate the accuracy of the predicted results, the mean absolute percentage error (MAPE) is calculated as follows:
\begin{equation}
\mathrm{MAPE}=\frac{100\%}{N_e}\sum^{N_e}_{m=1}|\frac{y_m-\hat{y}_m}{y_m}|,
\label{equ:mape}
\end{equation}where $N_e$ is the number of the input and output pairs in the testing dataset, $\hat{y}_m$ and $y_m$ are the predicted and observed values scaled back from logarithmic values, respectively. The smaller the MAPE, the higher the accuracy of the ANN model. The MAPE for the forecasting performance in Fig.~\ref{fig:forecast} is 7.6\%, a relatively low value in the load forecasting community for one-day ahead forecasting.

\vspace{-4pt}
\subsection{Attack Models}
Adversaries may have different skill levels, e.g., compromising meter readings, gaining the access to a business network, or hijacking industrial control systems~\cite{majumder2019cyber}, to apply their attack strategies on the forecasting data. Eventually, an adversary needs to alter the data, causing an inaccurate forecasting result. A variety of forms can be used to model different attacks, among which the scaling attack, ramping attack, and random attack~\cite{cui2019machine,yue2019descriptive,sridhar2014model} have often been adopted in the literature. In this paper, the three attack models are also utilized.

Note that, 1) the three attack models are somehow representative; many other attack templates, e.g., pulsing attack, or more sophisticated attacks, can be represented by one of these attacks via a specific configuration, or by a proper combination of these models; and 2) the AML-based load forecasting works not only for these three attack models but also for a wide range of other attacks; as the focus of this paper is to investigate the performance of AML-based load forecasting, the three models are used to serve as an example in the paper.

\begin{figure}[t]
\centering
\includegraphics[width=0.48\textwidth]{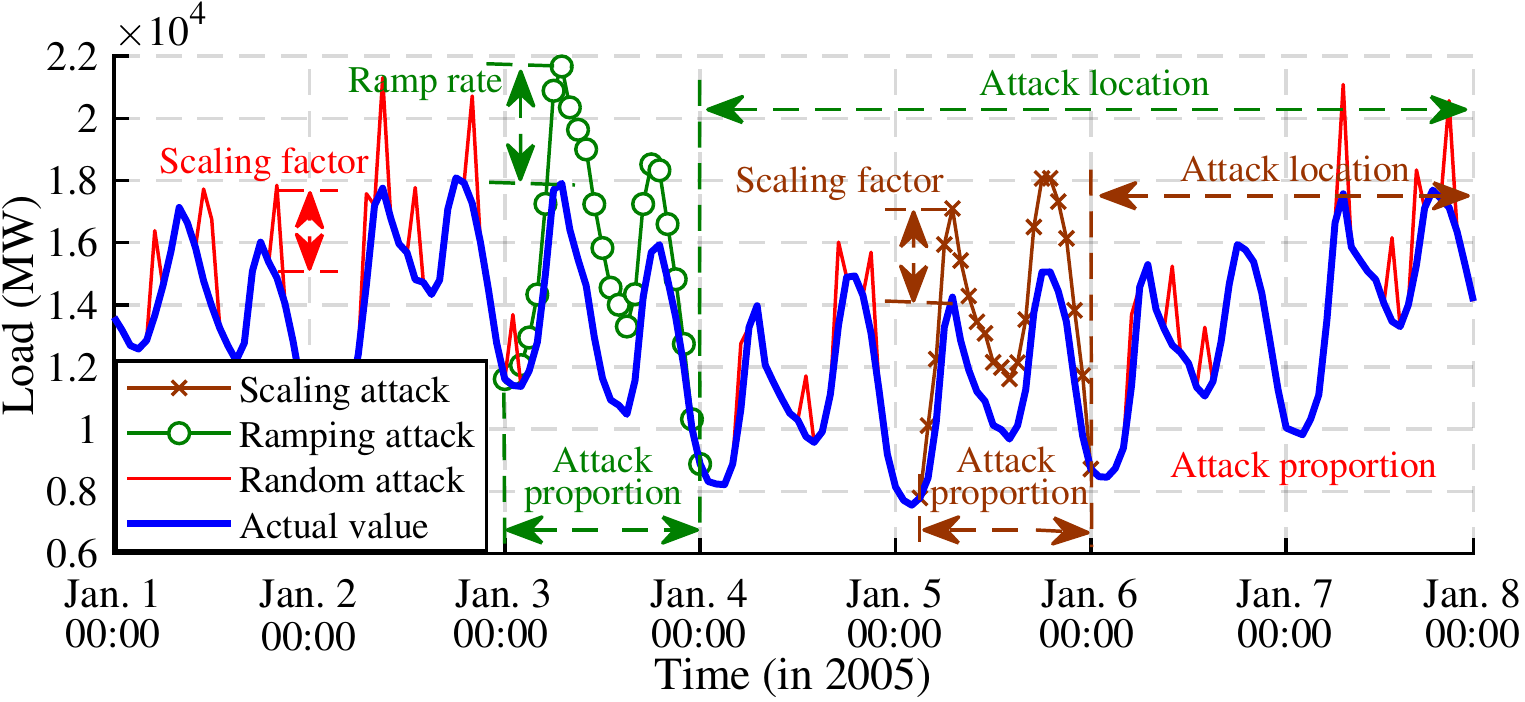}
\vspace{-4pt}
\caption{\label{fig:Attackmodel}An example of scaling, ramping and random attacks with the data from 1 January 2005 to 7 January 2005.}
\vspace{-8pt}
\end{figure}

An example of the three attack models is illustrated in Fig.~\ref{fig:Attackmodel}, where the data points are hourly loads obtained in one week from 1 January 2005 to 7 January 2005. The details of the three attack models are given below.

\subsubsection{Scaling Attack}
The adversary modifies the true measurements to higher or lower values during the entire duration of the attack via a constant scaling factor $\lambda_{se}$. Each data point $x_i$ is modified as
\begin{equation}
x_i^*=\lambda_{se}  x_i, \ i=n_s, n_s+1, ..., n_e,
\label{equ:scaling}
\end{equation}
where $x_i^*$ and $x_i$ are the compromised and actual data points at time $i$, respectively; and $n_s$ and $n_e$ are the start and end times of the attack, respectively. For an ANN model whose input consists of $N_{in}$ (i.e., 168 in this study) data points, $n_s$ and $n_e$ can be expressed as follows:
\begin{equation}
\begin{cases}
n_e=N_{in}-\gamma_{se}\\
n_s=n_e+1-p_{se}N_{in},
\label{equ:nsne}
\end{cases}
\end{equation}where $p_{se}$ is the attack proportion. It is the ratio of the number of attacked data points and the number of total data points. $\gamma_{se}$ is the attack location, which refers to the distance between the end time of the attack $n_e$ and the end of the total input data points.


\subsubsection{Ramping Attack}
The adversary modifies the true measurements gradually via a ramp rate $\lambda_{re}$ during the entire duration of the attack. Each data point $x_i$ is modified as
\begin{equation}
x_i^*=
\begin{cases}
 [1+   \lambda_{re}  (i-n_s)]x_i, \ i=n_s, n_s+1, ..., \lfloor\frac{n_s+n_e}{2}\rfloor \\
[1 +   \lambda_{re}  (n_e-i)]x_i, \ i=\lfloor\frac{n_s+n_e}{2}\rfloor+1, ..., n_e
\label{equ:ramping}
\end{cases}
\end{equation} 
where $\lfloor \frac{n_s+n_e}{2} \rfloor$ is the floored value of $\frac{n_s+n_e}{2}$~\cite{yue2019descriptive}. The definitions of $n_s$ and $n_e$ for the ramping attack are the same as those for the scaling attack (shown in (\ref{equ:nsne})). Here, we use $p_{re}$ and $\gamma_{re}$ to replace $p_{se}$ and $\gamma_{se}$, respectively, to better represent the parameters of ramping attack.

Note that the ramping attack ramps up the data at first and then ramps down the data, such that the attack is less noticeable than the scaling attack.

\subsubsection{Random Attack}
The adversary randomly selects a proportion $p_{de}$ of all the data points in the training dataset and modifies the selected true measurements via a constant scaling factor $\lambda_{de}$, as described below:
\begin{equation}
x_i^*=\lambda_{de}x_i.
\label{equ:random}
\end{equation}

As shown in Fig.~\ref{fig:Attackmodel}, the random attack has two parameters, namely the attack proportion $p_{de}$ and the scaling factor $\lambda_{de}$. The definition of $p_{de}$ is the same as those for $p_{se}$ and $p_{re}$.


\vspace{-8pt}
\subsection{Impact Analysis}



The impacts of the three attacks on the ANN load forecasting are evaluated in this subsection using the configurations shown in Table \ref{tab:para}. The configurations are set in a reasonable way based on the characteristics of the dataset. In this study, the input to the ANN model consists of 168 hourly data points in total. For the ramping attack, the maximum $p_{re}$ is set at $(168-\gamma_{re})/168$ to represent the proportion of the attacked data from the beginning of the week (represented as 168) to the attack location $\gamma_{re}$ (see Fig. \ref{fig:Attackmodel}). For the scaling attack, the maximum $p_{se}$ is set at 0.1428 such that one day of load data are attacked. It is reasonable to set the maximum $p_{se}$ at 0.1428 while setting the maximum $p_{de}$ at one, because the attacks on more recent data (namely, the data more closely to the end of the week in Fig. \ref{fig:Attackmodel}) have a much greater impact on the forecasting. $\gamma_{se}$ and $\gamma_{de}$ are selected among $\{0, 24, 48, 72\}$. These numbers are selected as an example in this study to represent different attack locations.



\begin{table}
  \caption{Configurations of Attack Parameters}
  \vspace{-4pt}
  \centering
\begin{tabular}{c c c c}
  \hline
   \multirow{2}{*}{\shortstack{Scaling Attack}} &$\lambda_{se}$ &$p_{se}$ &$\gamma_{se}$ \\
   &[0.4, 2] &[0, 0.1428] & \{0, 24, 48, 72\} \\
   \multirow{2}{*}{\shortstack{Ramping Attack}} &$\lambda_{re}$ &$p_{re}$ &$\gamma_{re}$ \\
   &[0, 1] &[0, (168-$\gamma_{re}$)/168] &\{0, 24, 48, 72\} \\
   \multirow{2}{*}{\shortstack{Random Attack}} &$\lambda_{de}$ &$p_{de}$ & \\
   &[0.4, 2] &[0, 1] & \\
  \hline
\end{tabular}
\label{tab:para}
\vspace{-12pt}
\end{table}

\begin{figure}[b]
\centering
\includegraphics[width=0.48\textwidth]{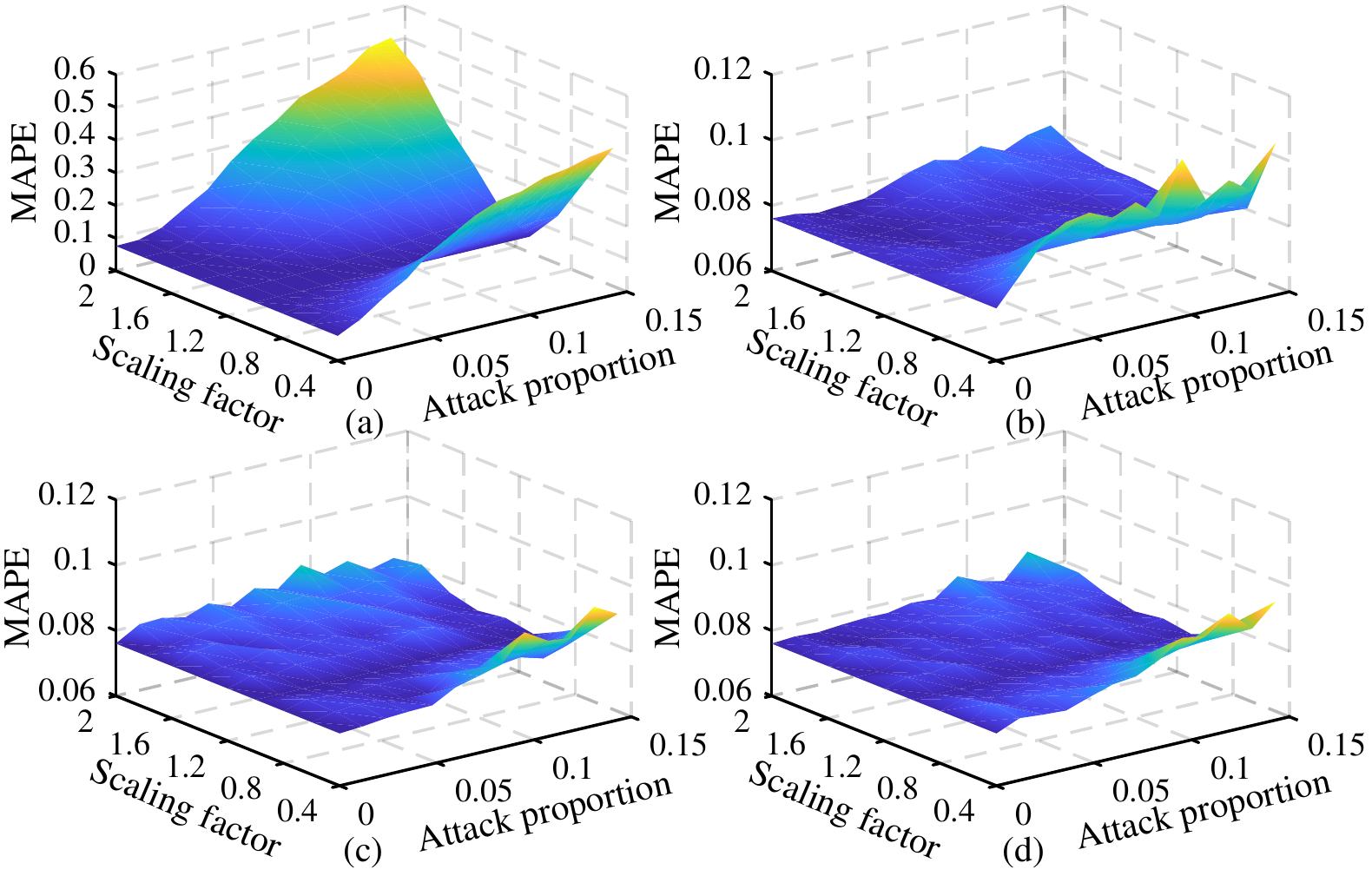}
\vspace{-8pt}
\caption{\label{fig:ImpactScaling}The performance of ANN load forecasting under the scaling attack. (a) $\gamma_{se}$ is 0. (b) $\gamma_{se}$ is 24. (c) $\gamma_{se}$ is 48. (d) $\gamma_{se}$ is 72.}
\end{figure}

\begin{figure}[ht]
\centering
\includegraphics[width=0.48\textwidth]{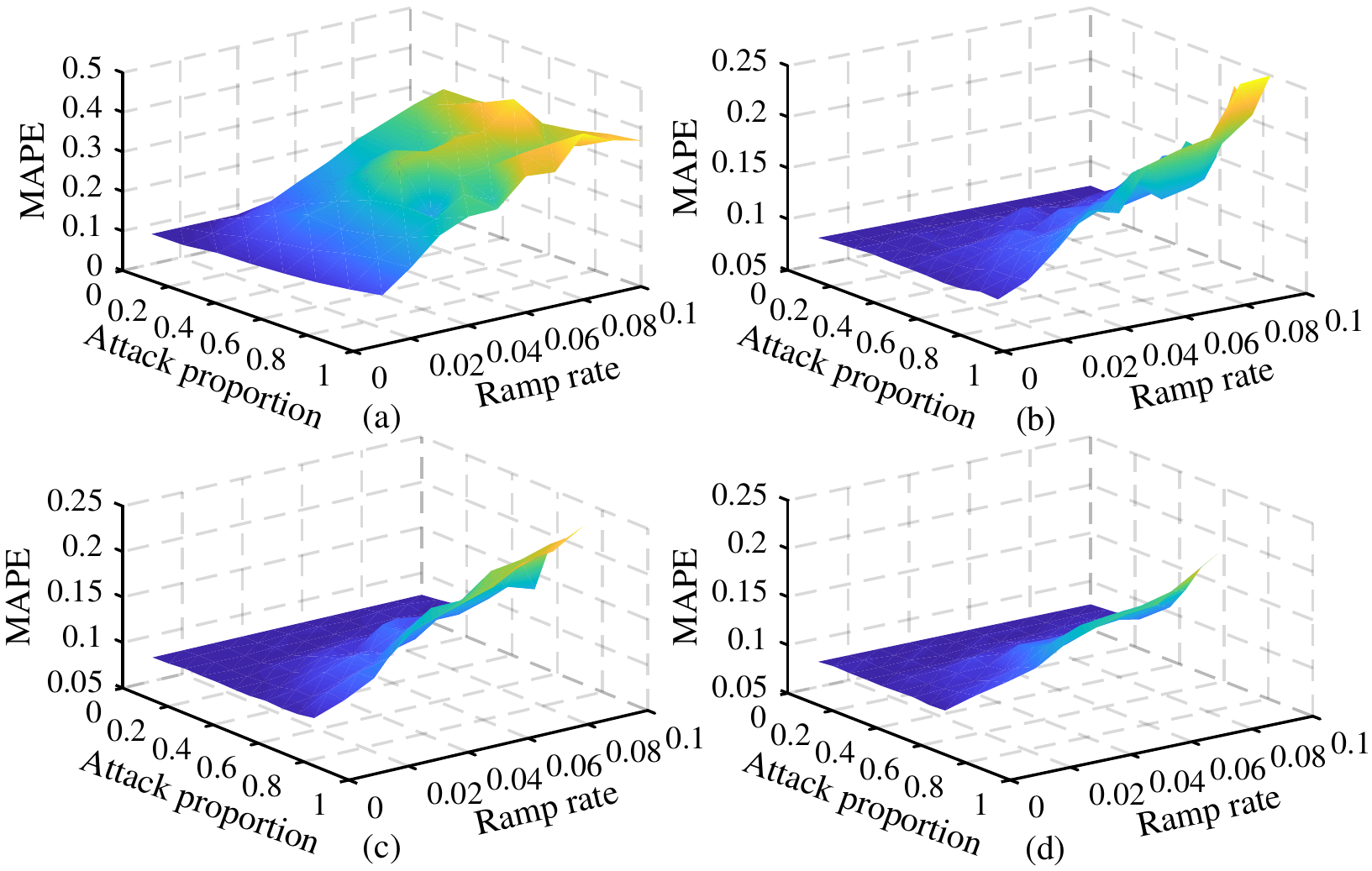}
\vspace{-8pt}
\caption{\label{fig:ImpactRamping}The performance of ANN load forecasting under the ramping attack. (a) $\gamma_r$ is 0. (b) $\gamma_r$ is 24. (c) $\gamma_r$ is 48. (d) $\gamma_r$ is 72.}
\end{figure}

\begin{figure}[ht]
\centering
\includegraphics[width=0.24\textwidth]{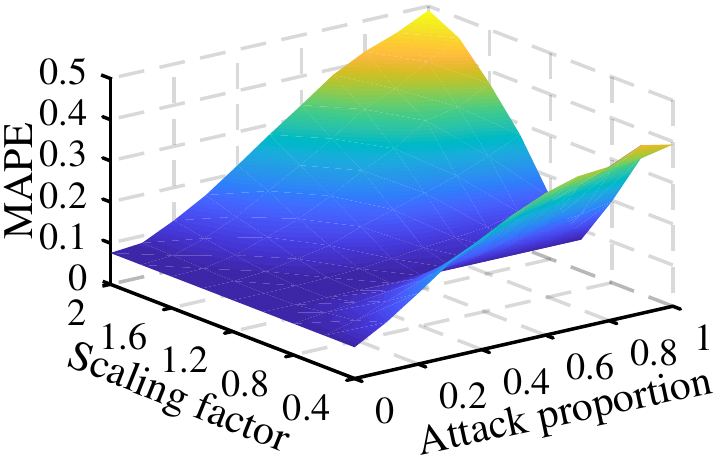}
\vspace{-6pt}
\caption{\label{fig:ImpactRandom}The performance of ANN load forecasting under the random attack.}
\vspace{-12pt}
\end{figure}

The performances of the ANN load forecasting model under the three attacks are illustrated in Figs. \ref{fig:ImpactScaling}-\ref{fig:ImpactRandom}, respectively. It can be seen that, 1) for each attack model, the impact can be severe under certain configurations, and this is why robust versions of forecasting are needed; and 2) the attacks on more recent data have much greater impact on the forecasting.

\vspace{-4pt}
\section{AML-based Load Forecasting}
\label{sec:aml}

This section describes the main issues specific to AML-based load forecasting, including 1) the generation of adversarial examples, 2) adversarial training process, 3) accelerated adversarial training, and 4) the framework.

\vspace{-8pt}
\subsection{Adversarial Examples}
In a traditional ANN, the model's input training data are clean, which renders the forecasting results sensitive to malicious data as illustrated in Figs. \ref{fig:ImpactScaling}-\ref{fig:ImpactRandom}. AML enhances robustness against attacks via an adversarial training such that the input training data are augmented with adversarial examples.

Most of the existing works on generating adversarial examples are focused on image recognition. An adversarial example $x^{adv}$ is generated such that it should have a small distance to the true input $x$, while at the same time resulting in a different output. Note that $x^{adv}$ and $x$ are vectors that contain multiple single values. Mathematically, $x^{adv}$ is defined as follows~\cite{cai2018curriculum}:
\begin{equation}
f_W(x^{adv})\neq z \wedge d(x, x^{adv})\leq \epsilon,
\label{equ:adv}
\end{equation}
where $W$ is the weight matrix of the neural network, $f_W(\cdot)$ represents the output of the network, $z$ is the actual value, $\epsilon$ is a given small value, and $d(\cdot,\cdot)$ represents the distance between two vectors, e.g., $x$ and $x^{adv}$ in (\ref{equ:adv}).

Generating an adversarial example is equivalent to finding an $x^{adv}$ that satisfies (\ref{equ:adv}). It is, however, difficult for attackers to directly solve (\ref{equ:adv}). One modification is to convert (\ref{equ:adv}) to the following optimization problem:
\begin{equation}
\underset{x^{adv}}{\mathrm{argmax}} \  L(f_W(x^{adv}), z)
\label{equ:obj}
\vspace{-8pt}
\end{equation}

\begin{equation}
\mathrm{s.t.} \ d(x, x^{adv}) \leq \epsilon,
\label{equ:constraints}
\end{equation}
where $L(\cdot, \cdot)$ is the squared error loss as defined in (\ref{equ:sel}). From (\ref{equ:obj}) and (\ref{equ:constraints}), an adversarial example $x^{adv}$ is generated based on the idea that once the squared error loss is maximized, it is more likely that the output of the network will be altered.

\vspace{-8pt}
\subsection{Adversarial Training}
Adversarial training aims to optimize the weight matrix $W$, which minimizes the upper bound of (\ref{equ:obj}) for all $x^{adv}$. Mathematically, the objective is expressed as
\begin{equation}
\underset{W}{\mathrm{argmin}} \underset{(x, z)\in D}{\mathrm{max}} \ \underset{d(x, x^{adv})\leq \epsilon}{\mathrm{max}} L(f_W(x^{adv}), z),
\label{equ:advtraining}
\end{equation}
where $D$ is the dataset, e.g., the training dataset. The idea of adversarial training is to solve (\ref{equ:advtraining}) by iteratively executing the following two steps~\cite{cai2018curriculum}: 1) with all given $x^{adv}$, find the optimal $W$ for the outer minimization problem, and 2) with the given $W$, find the worst-case adversarial example $x^{adv}$ in the dataset $D$ for the left inner maximization problem.

The standard SGD method is used to train the network by estimating the weight matrix $W$. Similarly to (\ref{equ:weight}), each weight $w$ is updated as follows:
\begin{equation}
w_{j+1}=w_j-\frac{\eta}{N_r}\sum_{m=1}^{N_r} \nabla_w L(f_W(x_{m,j}^{adv}), z_{m}),
\label{equ:sgd}
\end{equation}
where $x_{m,j}^{adv}$ is the adversarial example with regards to the $m^{th}$ input and output pair at the $j^{th}$ iteration.

\vspace{-8pt}
\subsection{Accelerated Adversarial Training}
In order to produce a different network output, the adversary uses (\ref{equ:obj}) and (\ref{equ:constraints}) to generate an adversarial example for an existing work such as image recognition. In the context of cyberattack-resilient load forecasting, however, the need to use the optimization to generate the adversarial examples is weak, because any change on the input vector for load forecasting directly leads to a different output. Further, according to (\ref{equ:advtraining}) and (\ref{equ:sgd}), the adversarial examples are re-generated at every iteration during the training stage, which inevitably makes the training process complicated and time-consuming. To make the process more efficient, we present an accelerated adversarial training in this paper. Two modifications are made as shown below:

\begin{itemize}
    \item Instead of re-generating the adversarial examples at each iteration, the presented training only generates the adversarial examples once (before the first iteration).
    \item To avoid the complex optimization process in (\ref{equ:obj}) and (\ref{equ:constraints}), the adversarial examples are generated simply based on the attack models, e.g., (\ref{equ:scaling})-(\ref{equ:random}). For instance, if the scaling attack is employed to generate adversarial examples, each input training data point is modified based on the attack parameters provided by (\ref{equ:scaling}).
\end{itemize}

This accelerated adversarial training still maintains robust performance against cyberattacks, as the input to the ANN model is augmented with the adversarial examples. But, at the same time, it becomes more efficient and easier to employ. The runtime of the algorithm to produce one output is only around 16 milliseconds in this study.

This accelerated adversarial training then estimates the weights iteratively according to (\ref{equ:sgd}) with constant adversarial examples $x_{m}^{adv}$ replacing the varying $x_{m,j}^{adv}$.

Note that, when generating the adversarial examples, the attack models are employed on the input training data. However, if there is an attack, the attack model will be on the input testing data. To distinguish between the attack models in the training dataset and those in the testing dataset, the attack parameters in the training dataset for the three attacks are represented as $\lambda_{sr}$, $p_{sr}$, $\gamma_{sr}$, $\lambda_{rr}$, $p_{rr}$, $\gamma_{rr}$, $\lambda_{dr}$ and $p_{dr}$.

In this paper, the adversarial training methods that use the scaling, ramping and random attack models are represented as ScalAdv, RampAdv and RandAdv, respectively.


\vspace{-8pt}
\subsection{Framework of AML-based Load Forecasting}
When employing the adversarial training approach for load forecasting, it was unclear how to set the attack model and its configuration for the adversarial training. Basically, there are two major concerns that are challenging but critical in this process:
\begin{itemize}
    \item Due to the unknown behavior of the adversary, both the attack model and its configuration are unknown to the defender. Thus, the adversarial training with a specific attack model and a given configuration may not perform well in reality under other attack scenarios.
    \item As the input training data are augmented with adversarial examples, a side-effect is that the accuracy of the load forecasting will become worse in normal cases when there is no attack.
\end{itemize}

To tackle the above challenges, this paper establishes a framework for the AML-based load forecasting to properly set the attack model and its configuration for the adversarial training. The procedures are formalized in Algorithm 1 and mainly consist of three stages.

In the first stage, a threshold $T_h$ is first determined by the defender. It restricts the error of load forecasting when there is no attack. For instance, if $T_h$ is 0.1, the MAPE of AML-based load forecasting should be lower than 0.1 using the clean testing data. Each attack model with each of its configurations is then employed individually to generate the adversarial examples and train the ANN model. The corresponding MAPEs are calculated using the trained ANN models and clean testing data. If a MAPE is lower than $T_h$, then the corresponding attack model and configuration are reserved; otherwise, this configuration is discarded.

In the second stage, every reserved attack model and each of its reserved configurations are employed to generate the adversarial examples and train the ANN model. The MAPEs are then calculated using the trained ANN models and the contaminated testing data under the same attack models with different configurations. For instance, if ScalAdv is employed with a specific configuration, each MAPE is calculated using the contaminated testing data under the scaling attack with each of its configurations. Then, for each reserved attack model, the configuration with the best overall performance (i.e., the overall lowest MAPE) under the same attack model with different configurations is selected.

In the last stage, among all the reserved attack models with the selected configurations, the one with the best overall performance when the testing data are attacked by all the attacks is selected.

\begin{algorithm}[t]
\SetAlgoLined
\textbf{Input:} Training and testing data, attack models, $T_h$\\
\textbf{Output:} Attack model, its configuration\\
 initialize $W$, $\eta$\;
 \For{every attack model}{
  \For{every configuration}{
   Generate $x^{adv}$ and train the ML model\;
   Calculate MAPE using the clean testing data\;
   \eIf{MAPE lower than $T_h$}{
    Reserve the configuration\;
   }
    {Discard the configuration\;}
  }
  \For{every reserved configuration}{
   Generate $x^{adv}$ and train the ML model\;
   Calculate MAPE using the testing data under the same attack model with different configurations\;
  }
  Select the configuration with the best performance\;
 }
Among all the reserved attacks, select the one that has the overall best performance when the testing data are attacked by all the attacks.
\caption{AML-based Load Forecasting}
\end{algorithm}

\section{Results}
\label{sec:result}
In this section, the comparison results are provided, which include 1) the AML approach's performance without attack, 2) an evaluation of different configurations, 3) a comparison of ScalAdv, RampAdv and RandAdv results under same attacks, 4) a comparison of the ScalAdv, RampAdv and RandAdv results under different attacks, and 5) the impact of $T_h$. The data used in this section are described in Section II Part A.

\subsection{AML Performance without Attack}
\begin{figure}[b]
\centering
\includegraphics[width=0.48\textwidth]{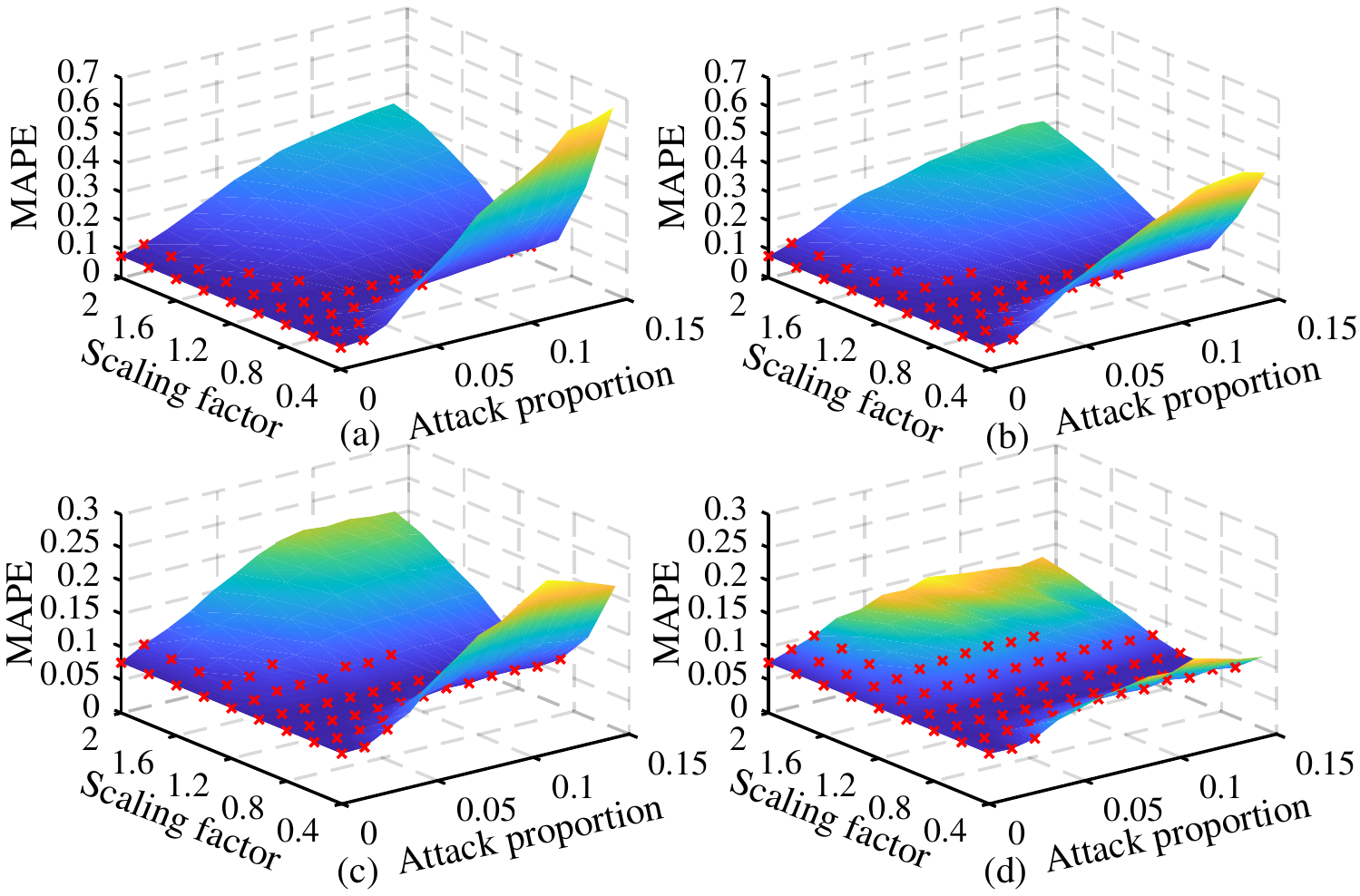}
\vspace{-8pt}
\caption{\label{fig:AdvScaNoAttack}The performance of ScalAdv-based load forecasting with different $\lambda_{sr}$, $p_{sr}$ and $\gamma_{sr}$ when there is no attack. (a) $\gamma_{sr}$ is 0. (b) $\gamma_{sr}$ is 6. (c) $\gamma_{sr}$ is 12. (d) $\gamma_{sr}$ is 18.}
\end{figure}

\begin{figure}[ht]
\centering
\includegraphics[width=0.48\textwidth]{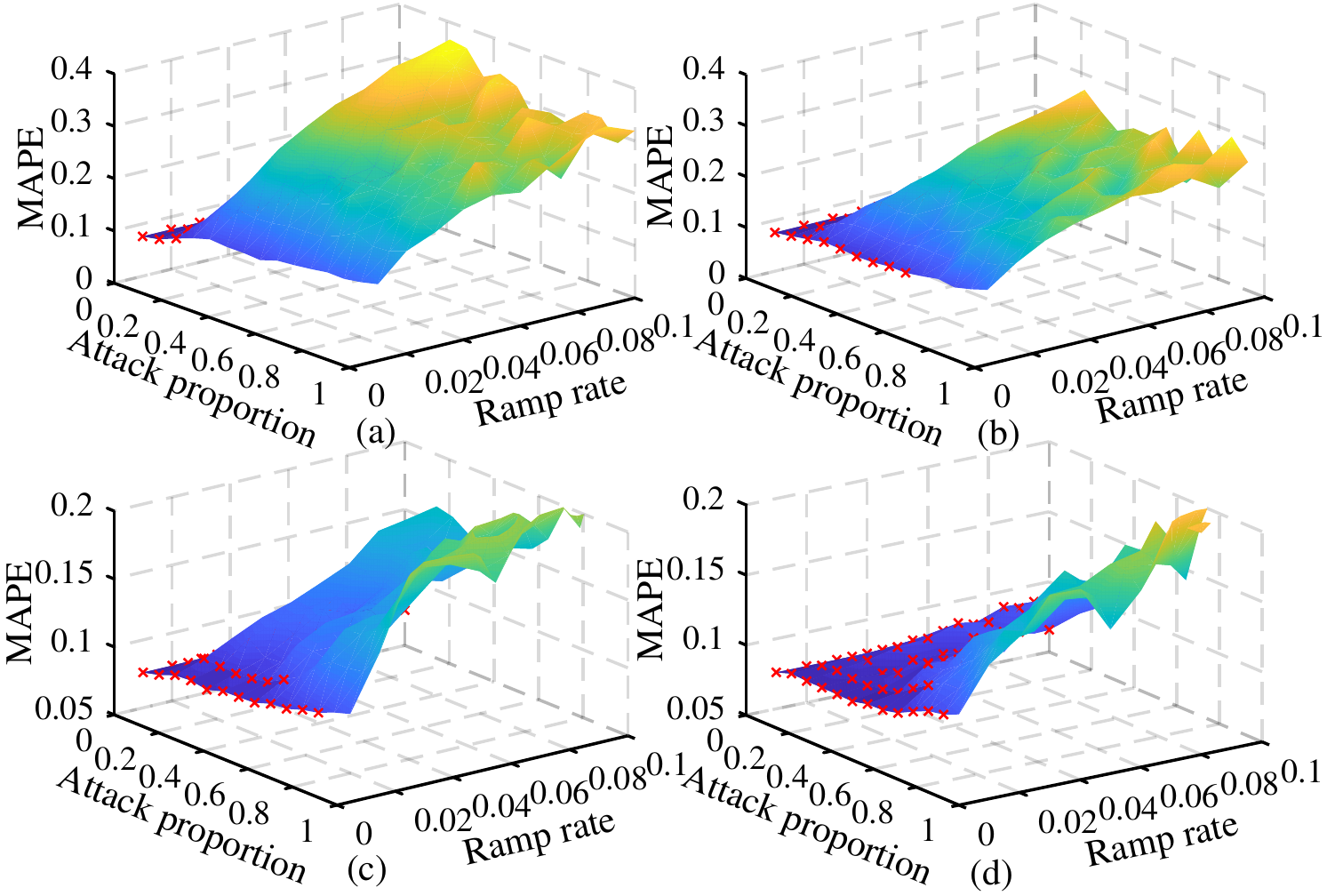}
\vspace{-8pt}
\caption{\label{fig:AdvRampNoAttack}The performance of RampAdv-based load forecasting with different $\lambda_{rr}$, $p_{rr}$ and $\gamma_{rr}$ when there is no attack. (a) $\gamma_{rr}$ is 0. (b) $\gamma_{rr}$ is 6. (c) $\gamma_{rr}$ is 12. (d) $\gamma_{rr}$ is 18.}
\end{figure}

\begin{figure}[ht]
\centering
\includegraphics[width=0.24\textwidth]{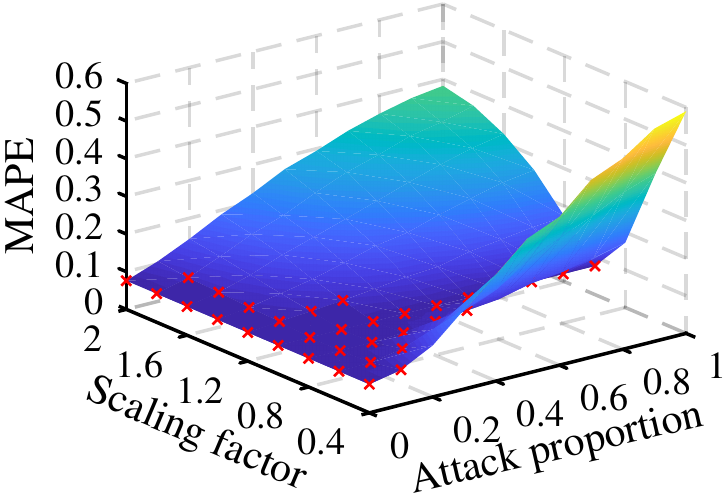}
\vspace{-8pt}
\caption{\label{fig:AdvRandNoAttack}The performance of RandAdv-based load forecasting with different $\lambda_{dr}$ and $p_{dr}$ when there is no attack.}
\vspace{-8pt}
\end{figure}

Figs. \ref{fig:AdvScaNoAttack}-\ref{fig:AdvRandNoAttack} give the performances of ScalAdv, RampAdv and RandAdv with different parameters when there is no attack. The configurations of the adversarial training's attack parameters are the same as those shown in Table \ref{tab:para}. The red points in Figs. \ref{fig:AdvScaNoAttack}-\ref{fig:AdvRandNoAttack} represent the MAPEs below 0.1.

It can be seen that 1) using different attack models and different configurations in the adversarial training both have different impacts when there is no attack, and 2) the impact can be undesirably large if the attack model and its configuration are not properly set. It is therefore critical to properly select the attack model and its configuration for a small impact when there is no attack. When the threshold $T_h$ is set at 0.1, all the configurations that correspond to the red points in Figs. \ref{fig:AdvScaNoAttack}-\ref{fig:AdvRandNoAttack} match the requirement and will be reserved.

\subsection{Evaluation of Different Configurations}
Among the reserved configurations, one configuration should be selected for each attack model in the adversarial training. To find the suitable one, the impact of each parameter is evaluated in this subsection.

\begin{figure}[t]
\centering
\includegraphics[width=0.5\textwidth]{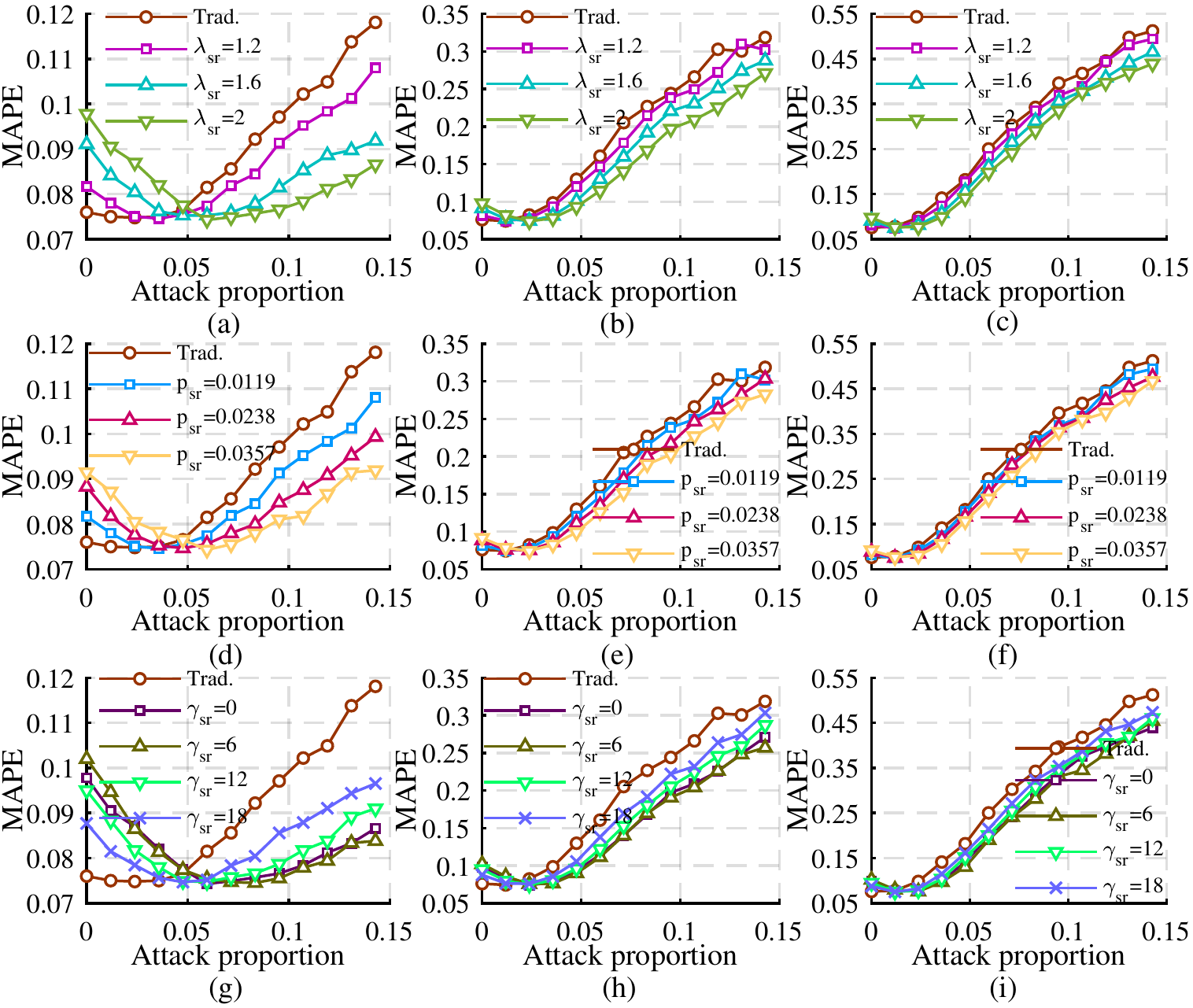}
\vspace{-20pt}
\caption{\label{fig:ScalAttackNine}The performance of ScalAdv-based load forecasting under the scaling attack with $\gamma_{se}$=0. (a)-(c) $\gamma_{sr}$ is 0, $p_{sr}$ is 0.0119, and $\lambda_{se}$ is 1.2, 1.6 and 2, respectively. (d)-(f) $\gamma_{sr}$ is 0, $\lambda_{sr}$ is 2, and $\lambda_{se}$ is 1.2, 1.6 and 2, respectively. (g)-(i) $\lambda_{sr}$ is 2, $p_{sr}$ is 0.0119, and $\lambda_{se}$ is 1.2, 1.6 and 2, respectively.}
\end{figure}

Table \ref{tab:MAPEScalAdv} gives the MAPEs of ScalAdv-based load forecasting when $\gamma_{sr}$ is 0 and there is no attack. The MAPEs in the shaded area are lower than $T_h$, i.e., 0.1. To evaluate the impact of $\lambda_{sr}$, the $\gamma_{sr}$ and $p_{sr}$ are fixed at 0 and 0.0119, respectively, and $\lambda_{sr}$ is tuned to 1.2, 1.6 and 2, respectively. The impact of $\lambda_{sr}$ under the scaling attack is illustrated in Fig. \ref{fig:ScalAttackNine} (a)-(c), where `Trad.' refers to the traditional load forecasting method that uses the normal ANN model. It can be seen that 1) when $\lambda_{se}$ is large, a larger $\lambda_{sr}$ will have lower MAPEs in most cases with different $p_{se}$; and 2) when $\lambda_{se}$ is small, even though a larger $\lambda_{sr}$ leads to higher MAPEs when $p_{se}$ is small, all those MAPEs are below $T_h$. Within the shaded area in Table \ref{tab:MAPEScalAdv}, a larger $\lambda_{sr}$ tends to perform better when under attacks.

The impact of $p_{sr}$ under the scaling attack is illustrated in Fig. \ref{fig:ScalAttackNine} (d)-(f). $\gamma_{sr}$ and $\lambda_{sr}$ are fixed at 0 and 1.2, respectively, and $p_{sr}$ is tuned to 0.0119, 0.0238 and 0.0357, respectively. It can be seen that, similarly to $\lambda_{sr}$, within the shaded area in Table \ref{tab:MAPEScalAdv}, a larger $p_{sr}$ performs better when under attacks.

Figs. \ref{fig:ScalAttackNine} (g)-(i) illustrate the impact of $\gamma_{sr}$ under the scaling attack. $\lambda_{sr}$ and $p_{sr}$ are fixed at 2 and 0.0119, respectively, and $\gamma_{sr}$ is tuned to 0, 6, 12 and 18, respectively. It can be seen that 1) a lower $\gamma_{sr}$ tends to have lower MAPEs in most cases with different $p_{se}$; and 2) even though a lower $\gamma_{sr}$ leads to higher MAPEs when $\lambda_{se}$ and $p_{se}$ are small, all those MAPEs are below $T_h$. Therefore, with a given $\lambda_{sr}$ and $p_{sr}$, a lower $\gamma_{sr}$ tends to perform better when under attacks.

Based on the above discussion, it can be observed that the three configurations marked in red in Table \ref{tab:MAPEScalAdv} tend to perform better under attacks than the others in the shaded area.

The selection of the parameters for RampAdv and RandAdv is not discussed here because similar observations were obtained for RampAdv and RandAdv. The desired configurations for RampAdv and RandAdv are set as the red ones in Table \ref{tab:MAPERampAdv} and Table \ref{tab:MAPERandAdv}, respectively.

\begin{table}
  \caption{MAPEs of ScalAdv-based Load Forecasting with $\gamma_{sr}$=0 under No Attack}
  \vspace{-4pt}
  \centering
\begin{tabular}{c c c c c c c}
  \hline
   \diagbox[linewidth=0pt,linecolor=white]{$\lambda_{sr}$}{$p_{sr}$} &0 &0.0119 &0.0238&0.0357&0.0476&0.0595 \\ \hline
   1.2&\cellcolor{blue!25}0.076&\cellcolor{blue!25}0.0817&\cellcolor{blue!25}0.0883&\cellcolor{blue!25}\color{red}0.0915&0.1038&0.1112\\
   1.4&\cellcolor{blue!25}0.076&\cellcolor{blue!25}0.0869&\cellcolor{blue!25}\color{red}0.0981&0.1122&0.1341&0.1512\\
   1.6&\cellcolor{blue!25}0.076&\cellcolor{blue!25}0.0911&0.1082&0.1331&0.1572&0.1831\\
   1.8&\cellcolor{blue!25}0.076&\cellcolor{blue!25}0.0956&0.1181&0.1531&0.1827&0.2208\\
   2.0&\cellcolor{blue!25}0.076&\cellcolor{blue!25}\color{red}0.0978&0.1301&0.168&0.2071&0.2464\\
  \hline
\end{tabular}
\label{tab:MAPEScalAdv}
\end{table}

\begin{table}
  \caption{MAPEs of RampAdv-based Load Forecasting with $\gamma_{rr}$=0 under No Attack}
  \vspace{-4pt}
  \centering
\begin{tabular}{c c c c c c c}
  \hline
   \diagbox[linewidth=0pt,linecolor=white]{$\lambda_{rr}$}{$p_{rr}$} &0 &0.0714 &0.1428&0.2142&0.2856&0.3570 \\ \hline
   0.01&\cellcolor{blue!25}0.076&\cellcolor{blue!25}0.0821&\cellcolor{blue!25}\color{red}0.095&0.1071&0.1165&0.1136\\
   0.02& \cellcolor{blue!25}0.076& \cellcolor{blue!25}0.0881 &0.1171& 0.1501&0.159&0.155\\
   0.03& \cellcolor{blue!25}0.076& \cellcolor{blue!25}0.093& 0.149& 0.1912& 0.2122&0.2064\\
   0.04& \cellcolor{blue!25}0.076& \cellcolor{blue!25}\color{red}0.098& 0.1752& 0.2371& 0.2455&0.2299\\
   0.05& \cellcolor{blue!25}0.076& 0.1054& 0.1961& 0.2682& 0.2771&0.268\\
  \hline
\end{tabular}
\label{tab:MAPERampAdv}
\end{table}

\begin{table}[ht]
  \caption{MAPEs of RandAdv-based Load Forecasting under No Attack}
  \vspace{-4pt}
  \centering
\begin{tabular}{c c c c c c c}
  \hline
   \diagbox[linewidth=0pt,linecolor=white]{$\lambda_{dr}$}{$p_{dr}$} &0 &0.1 &0.2&0.3&0.4&0.5 \\ \hline
   1.2&\cellcolor{blue!25}0.076&\cellcolor{blue!25}0.084&\cellcolor{blue!25}0.0901&\cellcolor{blue!25}\color{red}0.0978&0.1074&0.118\\
   1.4&\cellcolor{blue!25}0.076&\cellcolor{blue!25}0.087&0.1031&0.1202&0.1423&0.1665\\
   1.6&\cellcolor{blue!25}0.076&\cellcolor{blue!25}0.0931&0.118&0.15&0.1762&0.2031\\
   1.8&\cellcolor{blue!25}0.076&\cellcolor{blue!25}\color{red}0.0972&0.1324&0.1705&0.1997&0.238\\
   2.0&\cellcolor{blue!25}0.076&0.1067&0.1452&0.1887&0.2244&0.2667\\
  \hline
\end{tabular}
\label{tab:MAPERandAdv}
\end{table}

\subsection{AML Performance under Same Attacks}
In this subsection, the performances of the selected configurations in the previous subsection are compared under the same attack models. For the sake of convenience, the ScalAdv using the three configurations with MAPEs of 0.0915, 0.0981 and 0.0978 in Table \ref{tab:MAPEScalAdv} are represented as ScalAdv1, ScalAdv2 and ScalAdv3, respectively. Similarly, the RampAdv using the configurations with MAPEs of 0.095 and 0.098 in Table \ref{tab:MAPERampAdv} are represented as RampAdv1 and RampAdv2, respectively, and the RandAdv using the configurations with MAPEs of 0.0978 and 0.0972 in Table \ref{tab:MAPERandAdv} are represented as RandAdv1 and RandAdv2, respectively.

\begin{figure}[t]
\centering
\includegraphics[width=0.5\textwidth]{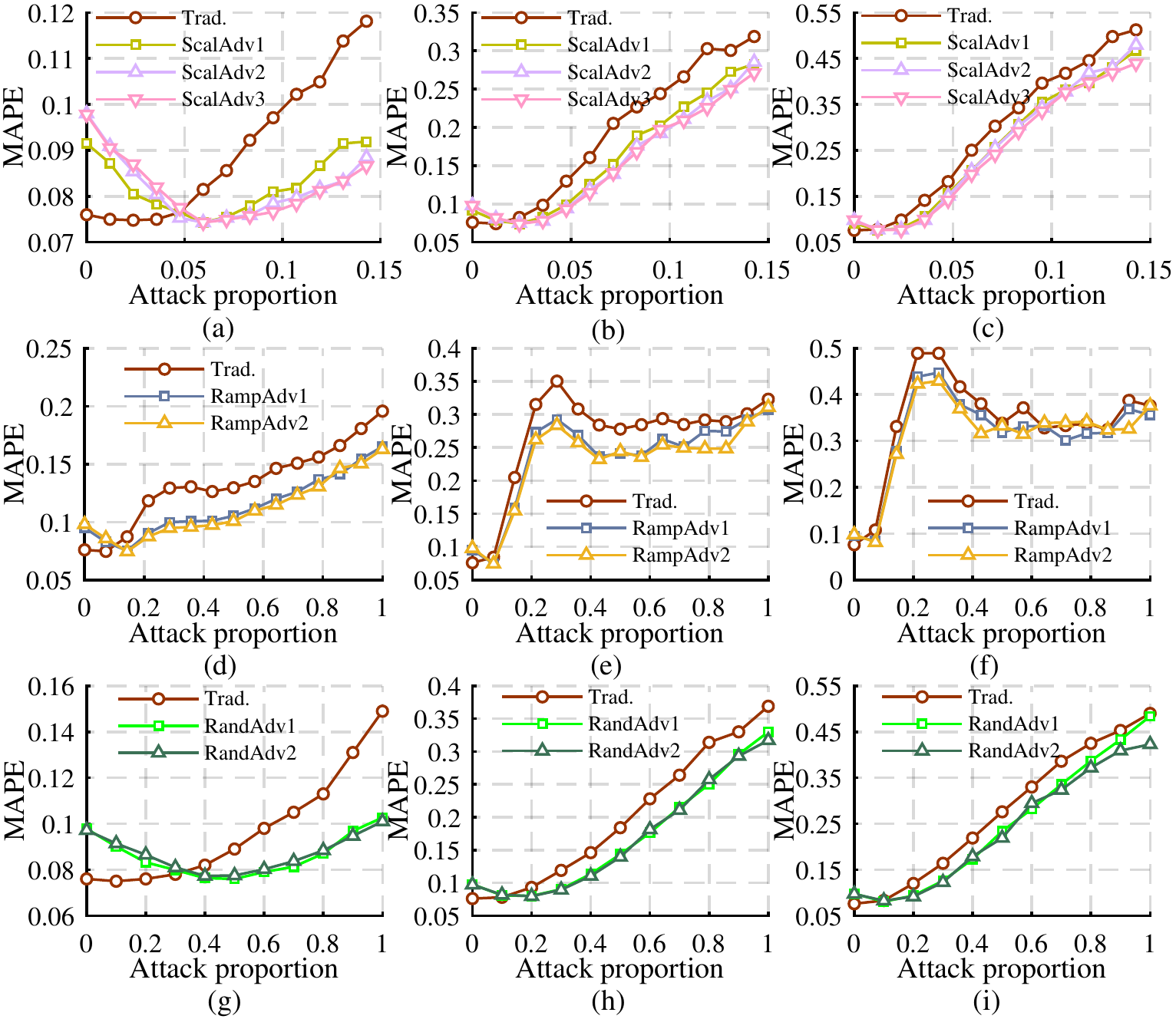}
\vspace{-20pt}
\caption{\label{fig:ScalRampRandAdvSameAttack}Comparison results of ScalAdv, RampAdv and RandAdv under same attack models. (a) Under scaling attack, $\lambda_{se}$ is 1.2. (b) Under scaling attack, $\lambda_{se}$ is 1.6. (c) Under scaling attack, $\lambda_{se}$ is 2. (d) Under ramping attack, $\lambda_{re}$ is 0.02. (e) Under ramping attack, $\lambda_{re}$ is 0.06. (f) Under ramping attack, $\lambda_{re}$ is 0.1. (g) Under random attack, $\lambda_{de}$ is 1.2. (h) Under random attack, $\lambda_{de}$ is 1.6. (i) Under random attack, $\lambda_{de}$ is 2.}
\vspace{-8pt}
\end{figure}

Figs. \ref{fig:ScalRampRandAdvSameAttack} (a)-(c) compare the results derived from ScalAdv1, ScalAdv2 and ScalAdv3 under the scaling attack; Figs. \ref{fig:ScalRampRandAdvSameAttack} (d)-(f) compare the results derived from RampAdv1 and RampAdv2 under the ramping attack; and Figs. \ref{fig:ScalRampRandAdvSameAttack} (g)-(i) compare the results derived from RandAdv1 and RandAdv2 under the random attack. For the scaling and ramping attacks, $\gamma_{se}$ and $\gamma_{re}$ are both fixed at zero. It can be seen that the selected configurations perform similarly under the same attack models. In this study, ScalAdv3, RampAdv2 and RandAdv2 are selected for ScalAdv, RampAdv and RandAdv, respectively, and their performances under different attack models will be compared in the next subsection.

\vspace{-4pt}
\subsection{AML Performance under Different Attacks}
\begin{figure}[t]
\centering
\includegraphics[width=0.5\textwidth]{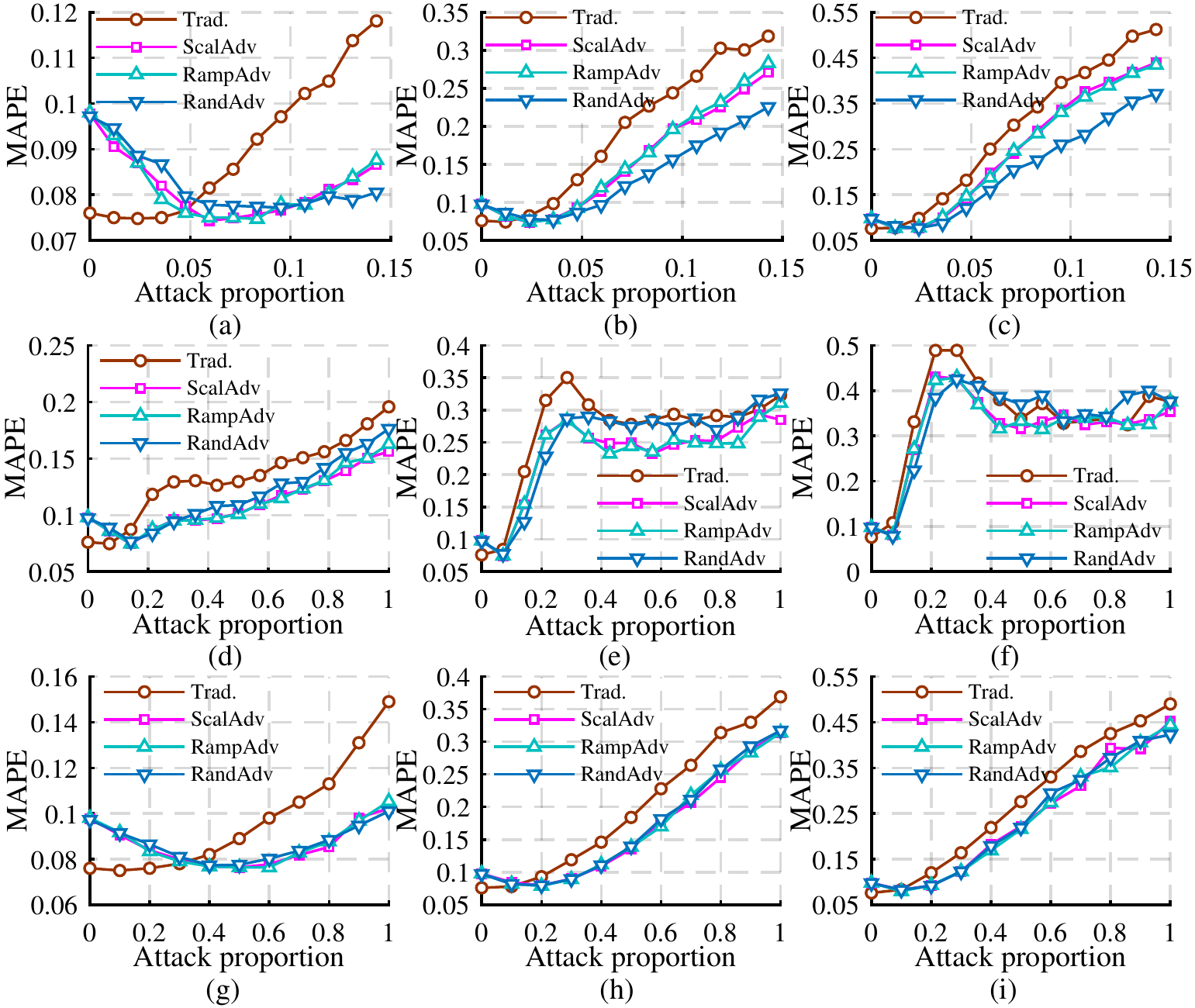}
\vspace{-20pt}
\caption{\label{fig:ScalRampRandAdv}Comparison results of ScalAdv, RampAdv and RandAdv under different attack models. (a) Under scaling attack, $\lambda_{se}$ is 1.2. (b) Under scaling attack, $\lambda_{se}$ is 1.6. (c) Under scaling attack, $\lambda_{se}$ is 2. (d) Under ramping attack, $\lambda_{re}$ is 0.02. (e) Under ramping attack, $\lambda_{re}$ is 0.06. (f) Under ramping attack, $\lambda_{re}$ is 0.1. (g) Under random attack, $\lambda_{de}$ is 1.2. (h) Under random attack, $\lambda_{de}$ is 1.6. (i) Under random attack, $\lambda_{de}$ is 2.}
\vspace{-8pt}
\end{figure}

This subsection investigates the AML approach's performance under other attacks. Figs. \ref{fig:ScalRampRandAdv} (a)-(c) compare the results of ScalAdv, RampAdv and RandAdv under the scaling attack; Figs. \ref{fig:ScalRampRandAdv} (d)-(f) compare the results of ScalAdv, RampAdv and RandAdv under the ramping attack; and Figs. \ref{fig:ScalRampRandAdv} (g)-(i) compare the results of ScalAdv, RampAdv and RandAdv under the random attack. Note that the configurations for ScalAdv, RampAdv and RandAdv here are the same as those for the previous section's ScalAdv3, RampAdv2 and RandAdv2, respectively.

It can be seen that 1) ScalAdv, RampAdv and RandAdv all have more robust performances than the traditional ANN even under other attacks, which validates the effectiveness and superiority of the AML-based methods in tackling the attacker's unknown behaviors; and 2) as long as the MAPE when there is no attack is fixed, the performances of ScalAdv, RampAdv and RandAdv are similar with each other, which is reasonable, because the total changes on the input training data for different AML approaches become similar.

Note that if other attack models, e.g., a combination of different attack models, are employed by the adversary, the AML-based load forecasting will have similar performance with those in Fig.~\ref{fig:ScalRampRandAdv}. That is, the performance of the AML-based approach with a given threshold would perform better than the traditional ANN method when the attack is strong, but slightly worse than the traditional ANN method when the attack is weak or when there is no attack.

\vspace{-4pt}
\subsection{Impact of $T_h$}
\begin{figure}[t]
\centering
\includegraphics[width=0.5\textwidth]{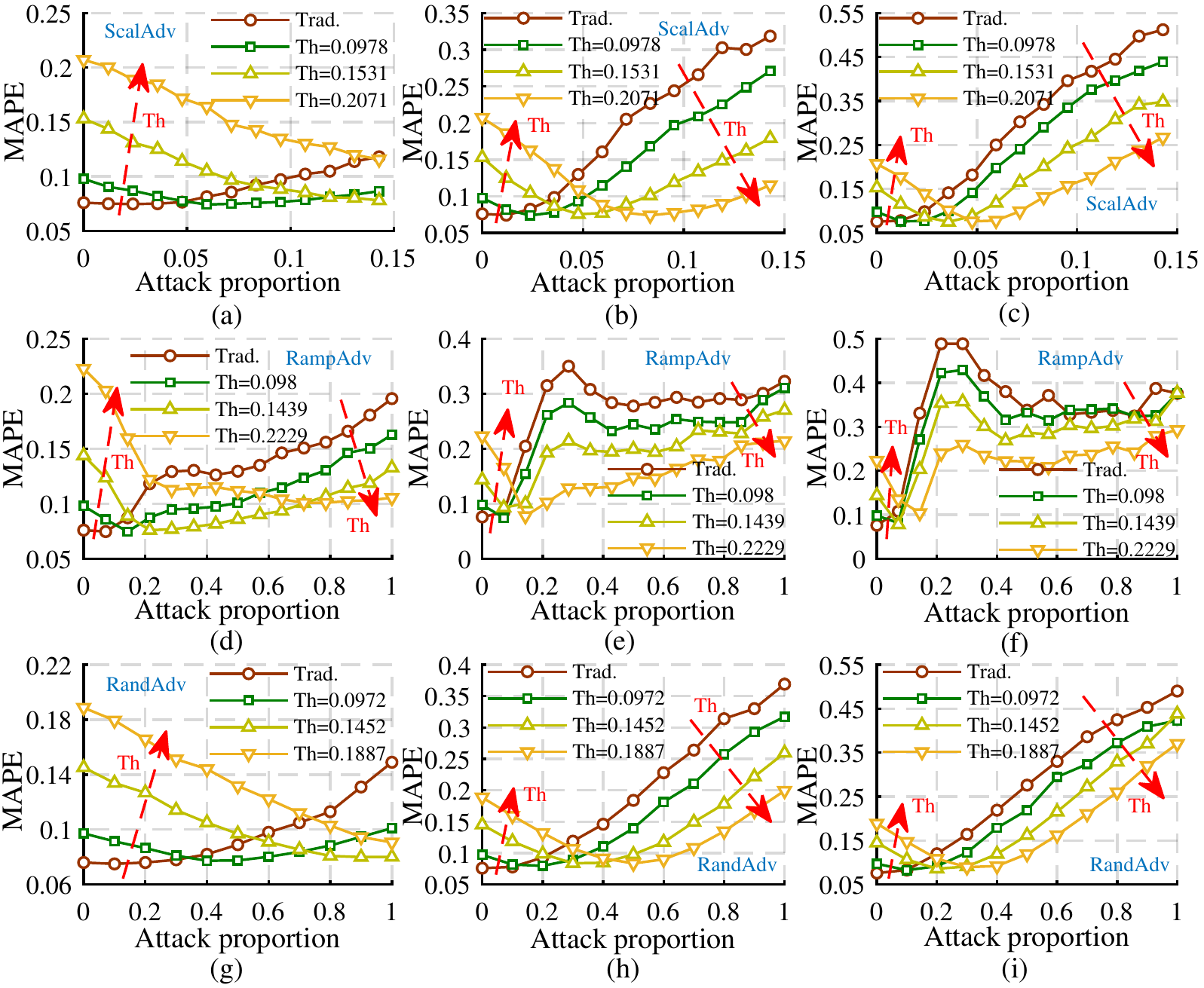}
\vspace{-20pt}
\caption{\label{fig:ThreCompar}Comparison results of ScalAdv, RampAdv and RandAdv with different $T_h$s under same attack models. (a) Under scaling attack, $\lambda_{se}$ is 1.2. (b) Under scaling attack, $\lambda_{se}$ is 1.6. (c) Under scaling attack, $\lambda_{se}$ is 2. (d) Under ramping attack, $\lambda_{re}$ is 0.02. (e) Under ramping attack, $\lambda_{re}$ is 0.06. (f) Under ramping attack, $\lambda_{re}$ is 0.1. (g) Under random attack, $\lambda_{de}$ is 1.2. (h) Under random attack, $\lambda_{de}$ is 1.6. (i) Under random attack, $\lambda_{de}$ is 2.}
\end{figure}

The impact of $T_h$ is evaluated in this subsection. Figs. \ref{fig:ThreCompar} (a)-(c) compare the results of ScalAdv with different $T_h$s under the scaling attack. The configuration of ScalAdv with each $T_h$, i.e., 0.0978, 0.1531 or 0.2071, is selected based on the MAPEs using ScalAdv when there is no attack as shown in Table \ref{tab:MAPEScalAdv}. For instance, when $\gamma_{sr}$=0, $\lambda_{sr}$=2 and $p_{sr}$=0.0476, the MAPE of load forecasting under no attack is 0.2071 (see Table \ref{tab:MAPEScalAdv}), which is one of the $T_h$s here.

Similarly, Figs. \ref{fig:ThreCompar} (d)-(f) compare the results of RampAdv with different $T_h$s under the ramping attack, and the configuration for each $T_h$, i.e., 0.098, 0.1439 or 0.2229, is selected based on the MAPEs using RampAdv under no attack. Figs. \ref{fig:ThreCompar} (g)-(i) compare the results of RandAdv with different $T_h$s under the random attack, and the configuration for each $T_h$, i.e., 0.0972, 0.1452 or 0.1887, is selected based on the MAPEs using RandAdv under no attack.

It can be observed that:
\begin{itemize}
    \item With a given $T_h$, the AML-based load forecasting performs particularly well when the attack is strong, i.e., the attack proportion and scaling factor/ramp rate are large.
    \item For each ScalAdv, RampAdv and RandAdv, a larger $T_h$ tends to have better performance when the attack is strong; but at the same time, the performance is worse when the attack is small. It is a trade-off between enhancing the robustness against cyberattacks and maintaining the forecasting accuracy when there is no attack.
\end{itemize}

In practice, the clean dataset is often obtained by using anomaly detection methods, which identify the rare observations such as outliers and abnormal patterns. In the future, this AML-based approach can be properly combined with other approaches, such that it is switched on only when alarms are raised by the anomaly detection methods.

\section{Conclusion}
\label{sec:conclusion}
This paper investigates the feasibility of applying AML for cyberattack-resilient load forecasting. While most existing works fail to tackle the attacker's unknown behaviors, the presented AML approach addresses this gap by developing an adversarial training that is able to employ a wide range of attack models. Future works could be done on applying this approach to forecasting hourly real-time system demand or peak load to provide more information for different electricity markets. This method could also be further improved by combining it with other approaches.

\ifCLASSOPTIONcaptionsoff
  \newpage
\fi

\bibliographystyle{IEEEtran}
\bibliography{IEEEabrv,mybibfile}

\end{document}